# Increasing the mobility and power-electronics figure of merit of AlGaN with atomically thin AlN/GaN digital-alloy superlattices

Nick Pant[1,2], Woncheol Lee[3], Nocona Sanders[2], and Emmanouil Kioupakis[2]

1. Applied Physics Program, University of Michigan, Ann Arbor, MI, 48109, U.S.A.
2. Department of Materials Science and Engineering, University of Michigan, Ann Arbor, MI, 48109, U.S.A.
3. Department of Electrical Engineering and Computer Science, University of Michigan, Ann Arbor, MI, 48109, U.S.A.

Corresponding Author: nickpant@umich.edu

**Abstract**

Alloy scattering in random AlGaN alloys drastically reduces the electron mobility and therefore the power-electronics figure of merit. As a result, Al compositions greater than 75% are required to obtain even a two-fold increase of the Baliga figure of merit compared to GaN. However, beyond approximately 80% Al composition, donors in AlGaN undergo the DX transition which makes impurity doping increasingly more difficult. Moreover, the contact resistance increases exponentially with increasing Al content, and integration with dielectrics becomes difficult due to the upward shift of the conduction band. Atomically thin superlattices of AlN and GaN, also known as digital alloys, are known to grow experimentally under appropriate growth conditions. These chemically ordered nanostructures could offer significantly enhanced figure of merit compared to their random-alloy counterparts due to the absence of alloy scattering, as well as better integration with contact metals and dielectrics. In this work, we investigate the electronic structure and phonon-limited electron mobility of atomically thin AlN/GaN digital-alloy superlattices using first-principles calculations based on density-functional and many-body perturbation theory. The band gap of the atomically thin superlattices reaches 4.8 eV, and the in-plane (out-of-plane) mobility is 369 (452) cm$^2$ V$^{-1}$ s$^{-1}$. Using the modified Baliga figure of merit that accounts for the dopant ionization energy, we demonstrate that atomically thin AlN/GaN superlattices with a monolayer sublattice periodicity have the highest modified Baliga figure of merit among several technologically relevant ultra-wide band-gap materials, including random AlGaN, $\beta$-Ga$_2$O$_3$, cBN, and diamond.

Power electronics that will drive the future electrical grid, and electric rail and aviation infrastructure need semiconductors with ultra-wide band gaps, high carrier mobilities, and shallow dopants[1,2]. Semiconductors with ultra-wide band gaps can tolerate high electric fields without electrical breakdown due to impact ionization. High carrier mobility ensures that electrical transport is energy efficient and does not generate unnecessary heat. Finally, shallow dopants with low ionization energies are necessary to efficiently introduce free electrons that conduct electricity. Using predictive first-principles calculations[3], we propose atomically thin superlattices of AlN and GaN as candidate semiconductors that satisfy all three criteria for the active region of next-generation power electronics. These semiconductors have been experimentally demonstrated for use in light-emitting diodes and are compatible with existing industrial manufacturing processes.

The performance of semiconducting materials in power-electronics applications is quantified by the Baliga figure of merit (BFOM)[4] and its modified version that accounts for dopant ionization[5]. The BFOM quantifies conduction losses, and is given by the expression, $\text{BFOM} = \epsilon_s \mu F_{br}^3/4$, where $\epsilon_s$ is the dielectric constant, $\mu$ is the carrier mobility, and $F_{br}$ is the critical breakdown electric field, which scales superlinearly with the band gap[4]. The cubic dependence of the BFOM on the breakdown field has led to intense research efforts in developing ultra-wide-band-gap semiconductors for power electronics. In this work, we use the BFOM and its modified version that accounts for dopant ionization to quantify the performance of semiconductors. For lateral power devices, the lateral figure of merit is an alternative metric to quantify conduction losses, and is given by $\text{LFOM} = e n_s \mu F_{br}^2$, where $n_s$ is the sheet carrier density[6]. Since the LFOM and BFOM depend very similarly on the mobility and breakdown field, we use the BFOM in our analysis for simplicity, however our conclusions would hold equally well using the LFOM as well. Although many ultra-wide-band-gap semiconductors, *e.g.*, AlN, diamond, and cubic boron nitride, exhibit promising BFOM, their lack of shallow dopants has hampered their adoption. Therefore, the modified BFOM[5], which is the BFOM multiplied by the dopant ionization ratio, is a more useful quantity for evaluating the performance of ultra-wide-band-gap semiconductors for power-electronics applications.

GaN and AlGaN are some of the most promising materials for highly efficient power-electronic devices. GaN technology is the state of the art for low to moderate power applications[7,8], *e.g.*, phone chargers, electric cars, and photovoltaic inverters, due to its wide band gap of 3.5 eV[9], high electron mobility of 800-1600 cm$^2$ V$^{-1}$ s$^{-1}$[10,11], and availability of shallow dopants[12]. The (modified) BFOM approximately scales with the band gap to the sixth power, therefore a promising approach for improving the figure of merit of GaN is increasing its band gap by alloying it with aluminum. The alloy Al$_x$Ga$_{1-x}$N is a solid solution of GaN and AlN, and has a band gap that can be tuned from 3.5 eV ($x = 0$) to 6.3 eV ($x = 1$)[13,14].

However, AlGaN alloys face several challenges regarding their doping and conductivity. Despite two decades of intense research, the anticipated gain to the performance of AlGaN has not been fully realized because the electrical conductivity drops dramatically as the Al composition increases. Below ~85% Al composition, the conductivity is limited by alloy scattering, which occurs due to the random occupation of Al and Ga in the lattice[15]. At the most disordered compositions of 50-60% Al, the electron mobility reaches a minimum that is seven times smaller than the electron mobility of GaN[16]. Consequently, Al compositions of ~75% are required to obtain even a two-fold increase of the (modified) BFOM compared to GaN. Unfortunately, at compositions greater than ~80%, the conductivity decreases again due to the donor DX transition[17,18], which occurs when donors, *e.g.*, Si or Ge, preferentially occupy interstitial sites rather than substitutional sites. This causes the donor transition level to lie deep within the band gap, which makes doping highly inefficient. Consequently, the modified BFOM decreases exponentially beyond an Al composition of ~85%.

Electrons in atomically ordered compounds, such as superlattices, do not undergo alloy scattering. Therefore, superlattices could offer a viable route toward increasing the mobility and modified BFOM of AlGaN at an Al composition where impurity doping is efficient. Fortunately, atomically thin superlattices of alternating AlN and GaN layers have been demonstrated using common growth techniques, *e.g.*, molecular-beam epitaxy[19–22] and metalorganic-vapor-phase epitaxy[23–25]. In the limit of atomic sublattice thickness, such ordered digital alloys show significant promise for performance improvements in light-emitting diodes compared to conventional random AlGaN alloys[26–28]. In contrast to previous work, which explored increasing the alloy-scattering mobility of the two-dimensional electron gas at the GaN/AlGaN interface with the insertion of an ultra-thin AlN interlayer[29–31], we are interested in using atomically thin AlN/GaN digital-alloy superlattices as the active region for power electronics.

In this work, we use atomistic calculations based on density-functional theory, density-functional perturbation theory, and many-body perturbation theory to uncover the electronic and electron-transport properties of atomically thin AlN/GaN superlattices, periodically repeating along the *c*-axis. Such structures retain the ultra-wide band gap of AlGaN, while exhibiting an enhanced phonon-limited mobility that is 3-4x larger than the mobility of random AlGaN alloys due to the absence of alloy disorder. Most importantly, these favorable properties occur at an effective composition of 50%, where impurity doping is efficient and there is good integration with contact metals and dielectrics. As a result, the atomically thin superlattices have the highest modified BFOM of all known ultra-wide-band-gap semiconductors, and show great promise for high-performance power electronics.

We investigated atomically thin AlN/GaN superlattices with two different stacking periods along the *c*-axis: one monolayer of AlN by one monolayer of GaN (1ML) stacking and two monolayers of AlN by two monolayers of GaN (2ML) stacking. We also calculated the electron transport properties of GaN and AlN to interpolate the phonon-limited mobility of random alloys. To simulate pseudomorphic strain on AlN substrates, we lattice-matched each semiconductor to the basal plane of AlN using the experimental lattice constant, while allowing the atomic positions and *c*-axis length to relax. We separately investigated the relaxation of the ground-state crystal structures by minimizing the total energy with respect to the atomic coordinates, and requiring all forces to be less than $10^{-3}$ Ry/Bohr and the total energy to be converged within $10^{-4}$ Ry. We performed band structure and phonon calculations using Quantum Espresso[32] in the local-density approximation (LDA)[33]. We used norm-conserving pseudopotentials for the $3s^2p^1$ valence electrons of Al, $3d^{10}4s^2p^1$ valence electrons of Ga, and $2s^2p^3$ valence electrons of N. We used a plane-wave kinetic energy cutoff of 130 Ry, and a converged 8×8×4 (8×8×2) Monkhorst-Pack Brillouin-zone sampling grid for the self-consistent calculation of the 1ML (2ML) superlattice, GaN, and AlN. For the non-self-consistent calculation and the phonon calculation, we used a coarse 8×8×8 (8×8×4) Monkhorst-Pack grid. We applied many-body quasiparticle corrections in the $G_0W_0$ approximation using BerkeleyGW to obtain accurate band gaps and effective masses[34,35]. To obtain the phonon-limited mobility, we iteratively solved the linearized Boltzmann transport equation using EPW[36]. This requires calculating the *ab initio* electron-phonon matrix elements from density-functional perturbation theory, which calculates the linear response of the Kohn-Sham potential to a collective atomic displacement through the linear response of the charge density. We included all interband and intraband scattering processes between thermally occupied electron $|n\mathbf{k}\rangle$ and phonon $|\nu\mathbf{q}\rangle$ states by integrating the electron-phonon matrix elements across the Brillouin zone. Additionally, we solved the alloy-scattering-limited mobility using an in-house code in the relaxation-time approximation, which is a valid approximation since alloy scattering is elastic and has no angular dependence. For details of our mobility calculations, we refer the reader to the Supplementary Material.

By performing structural-relaxation calculations, we found that the atomically thin superlattices of AlN and GaN are well suited for epitaxial growth on bulk AlN substrates. In contrast to traditional multi-quantum-well structures, whose critical thickness is independently limited by the bulk lattice constant of each sublattice layer, the critical thickness of atomically thin superlattices is determined by a single lattice constant that describes the entire superlattice structure. In Table 1, we list the relaxed in-plane lattice constants $a$ that we calculated for GaN, AlN, and the 1ML and 2ML AlN/GaN superlattices. Our calculated lattice constants for GaN and AlN are in good agreement with experiment, which we also list in Table 1. We additionally show the epitaxial strain $\epsilon$ of each material if coherently strained to the basal *c*-plane of AlN. The 1ML and 2ML superlattices exhibit a lattice mismatch of only 1.6% compared to AlN, which should enable thick pseudomorphic superlattice stacks on AlN substrates. To estimate the critical thickness $t_{crit}$, we can make use of the fact that the critical thickness scales inversely with the lattice mismatch, *i.e.*, $t_{crit} \sim 1/|\epsilon|$[37]. Recently, 30 nm thick pseudomorphic GaN layers in AlN/GaN/AlN double heterostructures were demonstrated on AlN substrates[38,39]. The lattice mismatch between GaN and AlN is two times greater than the lattice mismatch between the superlattices and AlN. Extrapolating from the experimentally demonstrated thickness of GaN on AlN, we roughly estimate superlattice stacks with thickness of ~60 nm to be experimentally feasible. Overall, we expect that thick stacks of atomically thin AlN/GaN superlattices can be grown on AlN substrates while being nearly free of misfit dislocations that are harmful for device operation.

**Table 1.** The relaxed in-plane lattice constants $a$ and the corresponding epitaxial strain $\epsilon$ if coherently grown on the basal *c*-plane of AlN. The experimental values are from Vurgaftman and Meyer[40]. The lattice constants of the superlattices are well described by Vegard's law.

| Material | $a$ [nm] (Theory) | $a$ [nm] (Experiment) | $\epsilon$ (Theory) | $\epsilon$ (Experiment) |
|---|---|---|---|---|
| GaN | 0.318 | 0.319 | -0.035 | -0.026 |
| 1ML AlN/GaN Superlattice | 0.312 | | -0.016 | |
| 2 ML AlN/GaN Superlattice | 0.312 | | -0.016 | |
| AlN | 0.307 | 0.311 | 0 | 0 |

Our band-structure results demonstrate that the atomically thin superlattices retain the ultra-wide band gap of random AlGaN alloys and exhibit dispersive conduction bands, indicating their promise for high-power devices. Figure 1 shows the quasiparticle band structure of the 1ML and 2ML AlN/GaN superlattices. For both structures, the band gap is direct at the $\Gamma$-point and energetically isolated from other valleys. We calculated band gaps of 4.6 eV and 4.3 eV for the 1ML and 2ML structures. We verified these values by comparing to previous calculations[28] that explicitly included the computationally expensive semicore Ga *$3s^2p^6$* electrons in addition to the *$3d^{10}$ $4s^2p^1$* valence electrons that we considered in the present work. The band gaps of the 1ML and 2ML structures, calculated with the semicore pseudopotentials, are 4.8 eV and 4.6 eV. These values are in good agreement with the band gaps calculated in the present work with valence pseudopotentials, and justify the choice of treating the *$3s^2p^6$* states as frozen core electrons. The band-gap results are summarized in Table 2, and show excellent agreement with optical measurements by Wu *et al*[22]. In Table 3, we list the basic *ab initio* electronic and electron-transport properties of GaN, AlN, and the atomically thin superlattices, namely the effective masses ($m^*$), the room-temperature electron mobility ($\mu$), the frequency of the highest longitudinal-optical (LO) mode ($\hbar\omega_{LO}$) and the static dielectric constant ($\epsilon_s$). As input to our mobility calculation, we use the $G_0W_0$-corrected eigenvalues. We also use the electron-phonon matrix elements calculated using density-functional perturbation theory at the LDA level, which is a valid approximation since LDA wave functions are nearly identical to $G_0W_0$ wave functions in common semiconductors[34], and therefore should give accurate electron-phonon matrix elements. Our mobility results agree with Monte-Carlo simulations[41] and experimental measurements[42] of the mobility of AlN to within 25%, which is typical for first-principles calculations[43]. The effective mass, frequency of the highest LO mode, and dielectric constants of the 1ML superlattice are close to linear interpolations of the end binary compounds. Therefore, as the sublattice thickness decreases, the atomically thin superlattices (approximately) approach the virtual-crystal limit.

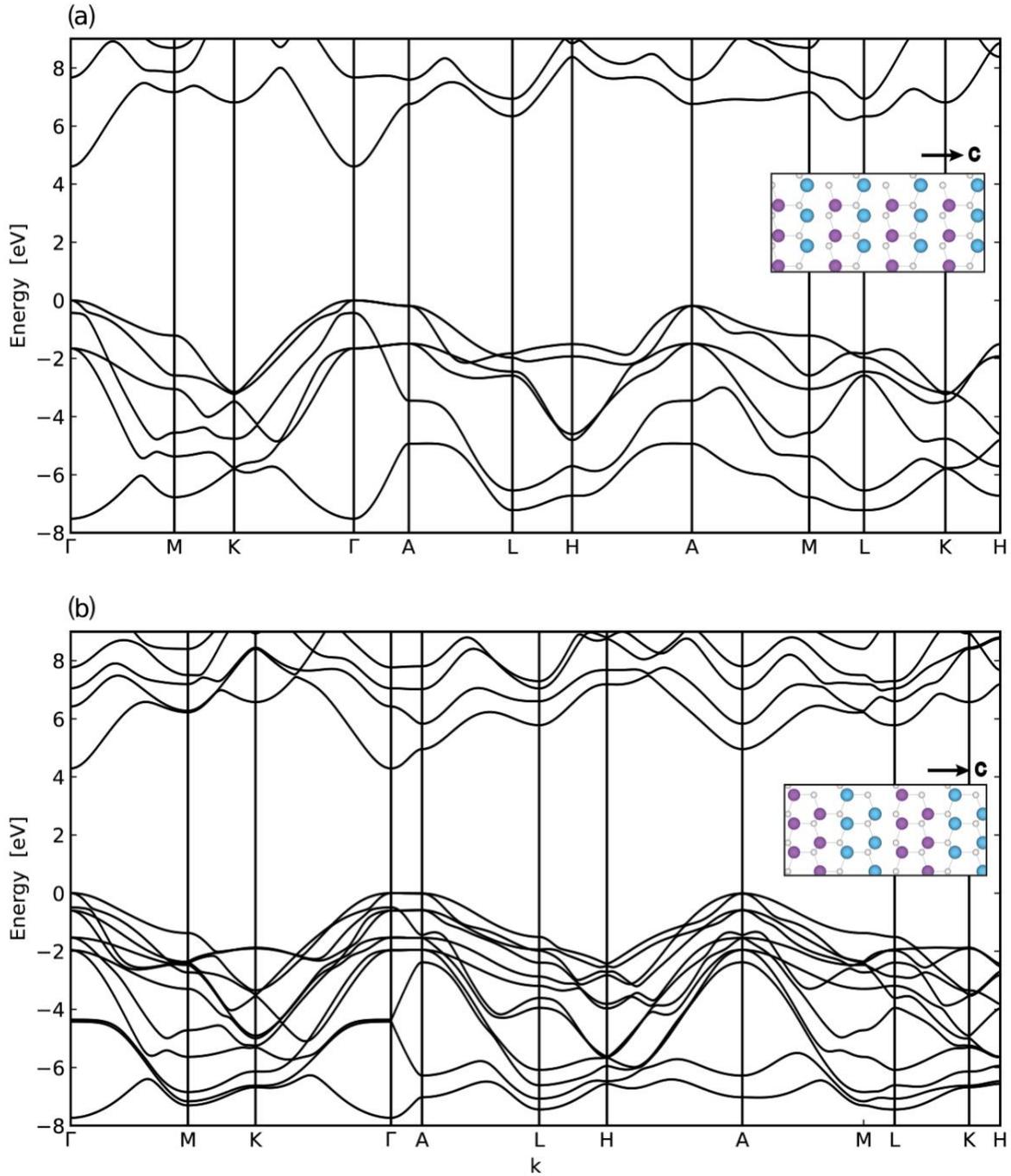

**Figure 1.** Quasiparticle band structure of (a) one-monolayer AlN / one-monolayer GaN superlattice and (b) two-monolayers AlN / two-monolayers GaN superlattice, periodically repeating along the *c*-axis. Both structures are pseudomorphically strained to AlN on the c-plane. The structural models for the superlattices are shown in the insets, with the wurtzite *c*-axis pointing to the right. The ultra-wide band gaps for both structures allow the materials to tolerate high electric fields without undergoing dielectric breakdown due to impact ionization.

**Table 2.** Theoretical quasiparticle band gaps (in eV) of atomically thin AlN/GaN superlattices. In this work, we treat the Ga $3s^2p^6$ electrons as frozen core states, and obtain good agreement with previous calculations that explicitly treat them in the valence[28]. The experimental optical gap, measured by Wu *et al.*[22], agrees with the theoretical predictions once excitonic effects are considered.

| Superlattice Stacking Period (AlN / GaN) | Theory (electronic, this work) | Theory (electronic, previous work) | Theory (optical, previous work) | Experiment (optical) |
|---|---|---|---|---|
| 1 ML / 1 ML | 4.6 | 4.8 | 4.7 | |
| 2 ML / 2 ML | 4.3 | 4.6 | 4.5 | |
| 1 ML / 2 ML | | 5.0 | 4.9 | 4.9 |

**Table 3.** Transport parameters (effective mass, room-temperature electron mobility, energy of the highest LO mode, and static dielectric constant) obtained from first-principles calculations. All materials are pseudomorphically lattice-matched to AlN on the c-plane, while the atoms and the *c*-axis length are allowed to relax.

| Material | $m^*_\perp/m_0$ | $m^*_\parallel/m_0$ | $\mu_\perp$ [cm² V⁻¹ s⁻¹] | $\mu_\parallel$ [cm² V⁻¹ s⁻¹] | $\hbar\omega_{LO}^{max}$ [meV] | $\epsilon_{s,\perp}/\epsilon_0$ | $\epsilon_{s,\parallel}/\epsilon_0$ |
|---|---|---|---|---|---|---|---|
| GaN | 0.25 | 0.21 | 430 | 721 | 93.7 | 9.3 | 10.6 |
| 1 ML AlN / 1 ML GaN Superlattice | 0.30 | 0.30 | 369 | 452 | 102.6 | 8.9 | 9.9 |
| 2 ML AlN / 2 ML GaN Superlattice | 0.31 | 0.33 | 210 | 212 | 102.3 | 8.8 | 10.1 |
| AlN | 0.32 | 0.33 | 373 | 283 | 114.4 | 8.0 | 9.6 |

Our electron transport calculations show that the mobility of atomically thin AlN/GaN superlattices is significantly higher than the mobility of random AlGaN alloys. We calculated the total mobility of random AlGaN alloys by combining the alloy-scattering-limited mobility of disordered AlGaN with the phonon-limited mobility of a virtual crystal, using Matthiessen's rule. In our previous work, we calculated the alloy-limited mobility of disordered AlGaN alloys whose lattice constants were fully relaxed[16]. To facilitate comparisons with the superlattices, which are pseudomorphically strained to AlN, we recalculated the mobility of random AlGaN alloys that are also pseudomorphically strained to AlN. We found that strain does not change the alloy scattering potential to within ~0.1 eV based on the conduction-band offset, but reduces the total mobility due to the increase of the effective mass. We calculated the virtual-crystal phonon-limited mobility by interpolating the mobility of GaN and AlN using an analytical model for piezoelectric scattering[44] that describes the functional dependence of the mobility on the effective mass, dielectric constant, and electromechanical coupling constant $K$, $\mu \propto \epsilon/(K^2 m^{3/2})$. We found that the total mobility of the alloy is, to first order, independent of the electron-phonon interpolation model used because of the dominance of alloy scattering. Compared to the alloy scattering potential of 1.8 eV, the scattering potential due to monolayer fluctuations is only ~0.1 eV, which is the energy difference between the conduction band of the 1ML and 2ML structures, evaluated by referencing their branch-point energies[45]. Therefore, we do not expect minor thickness fluctuations to significantly affect the mobility. In Figure 2, we compare the in-plane and out-of-plane room-temperature mobility of AlN/GaN superlattices and random AlGaN alloys, at a typical electron density of $10^{18}$ cm$^{-3}$. The superlattices exhibit enhanced mobility compared to random AlGaN due to the absence of alloy disorder. In particular, the 1ML superlattice exhibits an in-plane (out-of-plane) mobility that is 3.1× (3.8×) larger than the mobility of random Al$_{0.5}$Ga$_{0.5}$N. As mentioned earlier, the mobility of the 1ML superlattice is close to the virtual-crystal phonon-limited mobility. The difference in the mobility between the 1ML and 2ML superlattices can be qualitatively understood in terms of the fact that there are more phonon modes that can scatter electrons in the 2ML superlattice compared to the 1ML superlattice since there are eight atoms in the primitive cell of the 2ML superlattice compared to four atoms in the 1ML superlattice. Indeed, the thermally averaged relaxation time is approximately 30% larger in the 1ML superlattice than in the 2ML superlattice. The mobility calculated in the self-energy-relaxation-time approximation (SERTA) is 35-40% larger in the 1ML superlattice ($\mu_\perp$ = 209 cm$^2$/Vs, $\mu_\parallel$ = 207 cm$^2$/Vs) than in the 2ML superlattice ($\mu_\perp$ = 154 cm$^2$/Vs, $\mu_\parallel$ = 146 cm$^2$/Vs), which additionally reflects the increased effective mass in the 2ML structure. Interestingly, self-consistently solving the iterative Boltzmann transport equation increases the mobility of the 1ML superlattice by a factor of ~2, but the mobility of the 2ML superlattice increases only by ~40%, compared to the SERTA mobility. This suggests that the additional electron-phonon scattering pathways in the 2ML structure contribute more strongly to backward scattering, as opposed to forward scattering, than in the 1ML structure[46]. Nevertheless, electrons in the 2ML superlattice still exhibit a 1.6× greater mobility than in random Al$_{0.5}$Ga$_{0.5}$N. Therefore, replacing disordered AlGaN alloys with atomically thin superlattices is a viable solution for increasing the electron mobility.

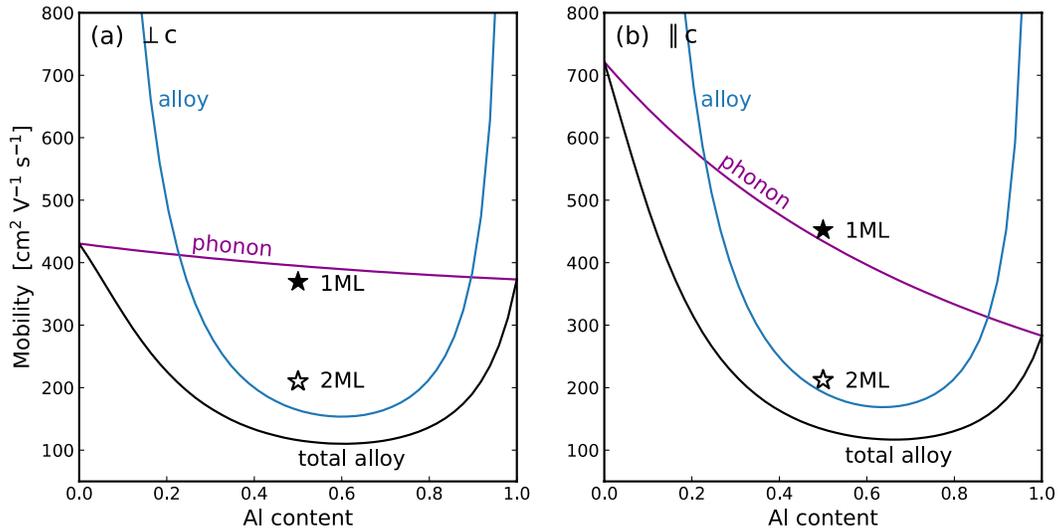

Figure 2. (a) In-plane ($\perp$c) and (b) out-of-plane ($\parallel$ c) mobility of atomically thin AlN/GaN superlattices compared to AlGaN alloys. The semiconductors are pseudomorphically strained to AlN on the c-plane. The mobility of the superlattice with one-monolayer (1ML) sublattice periodicity is indicated by the filled star, and the mobility of the two-monolayers (2ML) superlattice is indicated by the unfilled star. The black curve is the total mobility of a random alloy,

and the blue and purple curves show the alloy-scattering and phonon-scattering components, respectively. Both the in-plane and out-of-plane mobility of the superlattices exceed the mobility of random $Al_{0.5}Ga_{0.5}N$.

Our results show that the absence of alloy scattering in AlN/GaN superlattices increases the BFOM compared to both GaN and $Al_{0.5}Ga_{0.5}N$ alloys. To calculate the BFOM, we used the following formula to estimate the breakdown field, $F_{br} = 3.3 \text{ MV cm}^{-1} \times (\varepsilon_G/3.5)^2$, where 3.3 MV cm$^{-1}$ is the experimentally known breakdown field of GaN[47,48]. The model proposed by Higashiwaki *et al.*[48] slightly overestimates the breakdown field of ultra-wide-band-gap semiconductors compared to the model that we have used; however, we verified that both models support the conclusions of our work. In Figure SM1, we show that this simple phenomenological model properly describes the experimentally known breakdown fields in a wide range of semiconductors, including Si[48], GaAs[48], 4H-SiC[48,49], AlGaN[50–52], diamond[53], and $\beta$-$Ga_2O_3$[54], although these experiments are subject to large uncertainties. This model also agrees with theoretical calculations of the breakdown field using the Von Hippel criterion[55,56]. Accurate experimental measurements of the breakdown field do not yet exist for AlN and cBN. At a given (effective) composition, the breakdown field and dielectric constant in the superlattices are approximately equal to the breakdown field and dielectric constant in random AlGaN. However, the electron mobility is higher due to the absence of alloy scattering, thus the BFOM is also larger. In Figure 3, we show that the AlN/GaN superlattices exhibit greater BFOM than AlGaN alloys at an Al composition of 50% for both lateral and vertical transport. For reference, we have also shown the BFOM of *relaxed* GaN[10], *i.e.*, GaN that has not been pseudomorphically strained to AlN, which is the state-of-the-art for power electronics. The advantage of the superlattices is highlighted by the fact that Al compositions of ~75% is needed for random AlGaN alloys to obtain even a two-fold increase of its BFOM compared to GaN. For AlGaN alloys to be competitive with the 1ML superlattice, Al compositions greater than ~85% is needed, at which point dopants undergo the DX transition. At a much lower effective composition of 50%, the 1ML superlattice has a lateral (vertical) BFOM of 15 (18) MW/cm$^2$, and the 2ML superlattice has a lateral (vertical) BFOM of 6.5 (6.5) MW/cm$^2$, which are ~4× (~3×) and ~1.5× (~1.5×) times larger than in random $Al_{0.5}Ga_{0.5}N$.

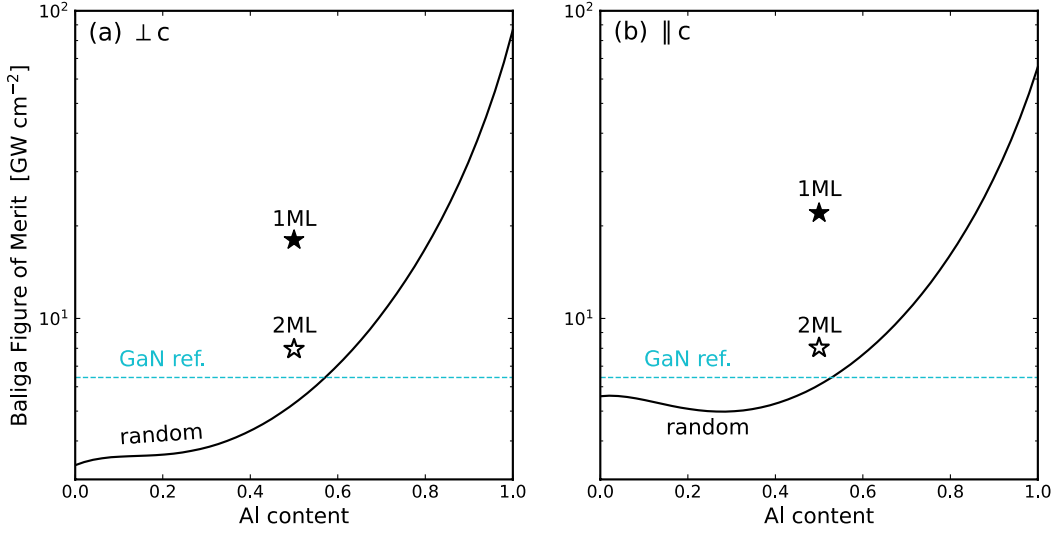

Figure 3. Baliga Figure of Merit for (a) lateral and (b) vertical transport in atomically thin AlN/GaN superlattices compared to AlGaN alloys. We assumed the breakdown field is related to the band gap according to, $F_{br} \propto \varepsilon_G^2$. The filled and unfilled stars show the BFOM of the one-monolayer (1ML) and two-monolayer (2ML) superlattices. The solid curve shows the BFOM of random AlGaN alloys. The dashed line shows the reference BFOM of relaxed GaN. All materials except the GaN reference are pseudomorphically lattice-matched to AlN.

We find that the modified BFOM, which accounts for dopant ionization, is higher in the atomically thin superlattices than in random AlGaN alloys throughout the entire composition range. We calculated the room-temperature dopant ionization ratio $\eta$ using the formula, $\eta = \left(1 + g \exp\left(\frac{\varepsilon_F - \varepsilon_D}{k_B T}\right)\right)^{-1}$, where $g$ is the degeneracy factor, $\varepsilon_F$ is the electron quasi-Fermi level, $\varepsilon_D$ is the dopant ionization energy, and $k_B T$ is the Boltzmann constant times the temperature. We assumed ultra-high purity of the materials, i.e., no charge compensation by impurities. We obtained $\varepsilon_D$ by empirically fitting a sigmoid function to the experimental ionization energies of Si in AlGaN, measured by Collazo et al.[18] (Figure SM2 in the supplementary material shows the dopant ionization energy of Si as a function of Al composition.) We numerically calculated the quasi-Fermi level for a fixed electron density of $10^{18}$ cm$^{-3}$ using the analytical 3D density-of-states expression. In Figure 4, we compare the modified BFOM of AlN/GaN superlattices and random AlGaN alloys. The modified BFOM of random AlGaN alloys reaches a maximum of 8.4 GW/cm$^2$ at an Al composition of 84%, very close to the DX transition. As we will show later in the text, this is higher than the modified BFOM of all known non-nitride semiconductors with experimentally demonstrated dopability. The modified BFOM of Al-rich AlGaN is exceeded only by the 1ML AlN/GaN superlattice, which exhibits a superior modified BFOM of 11.4 GW/cm$^2$ for vertical transport and 9.3 GW/cm$^2$ for lateral transport. Compared to random Al$_{0.5}$Ga$_{0.5}$N and GaN, the modified BFOM of the 1ML superlattice is ~300-400% greater. Although the modified BFOM of the 2ML superlattice is lower, it is still 65% greater than the modified BFOM of random Al$_{0.5}$Ga$_{0.5}$N and 95% greater than GaN. These results underscore the advantage of nitride semiconductors for high-performance and high-power applications.

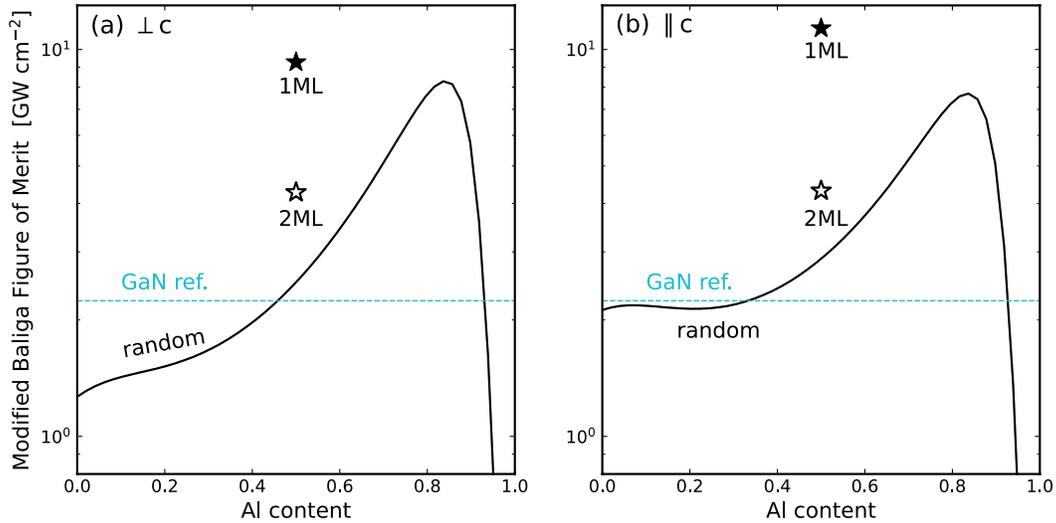

Figure 4. Modified Baliga Figure of Merit (BFOM) for (a) lateral transport and (b) vertical transport. The modified BFOM is the BFOM multiplied by the dopant ionization ratio, which we calculated using the dopant ionization energy measured by Collazo *et al*.[18] The vertical-transport modified BFOM of the 1ML superlattice is superior to random AlGaN throughout its composition range. Compared to the current state-of-the-art GaN technology (blue line), AlN/GaN superlattices offer performance improvements of up to 400%.

In addition to their improved mobilities and figure of merit, the atomically thin superlattices offer lower specific contact resistance to metals and better integration with dielectrics compared to Al-rich AlGaN alloys. An additional consideration in this comparison, which is not reflected in our work, is the experimental fact that random alloys are easier to grow than atomically thin superlattices. We address this by highlighting the technological advantages that the superlattices offer compared to random alloys, beyond what is reflected in the modified BFOM. We believe that these benefits warrant experimental effort on the growth and characterization of the superlattices. Although our calculations show that Al-rich random AlGaN alloys with Al composition below ~85% are promising in terms of their modified BFOM, their wider adoption has been hampered by the unfavorable position of their conduction band[2]. In particular, the large band offset between the conduction band of Al-rich AlGaN and the Fermi level of common ohmic-contact metals, *e.g.*, Ti- or V/Zr-based contacts, leads to a large barrier for electron tunneling between the metal and the semiconductor. This is problematic since the tunneling probability depends exponentially on the barrier height, *i.e.*, $P \propto \exp(-\sqrt{\phi_B}L)$, where $\phi_B$ is the energetic barrier and $L$ is the tunneling distance. Figure SM3 in the Supplementary Material shows the composition-dependent conduction-band position of random AlGaN alloys[14] and AlN/GaN superlattices, which we evaluated by referencing their branch point energies. The conduction band in the 1ML (2ML) superlattice is lower by 0.43 (0.57) eV than in $Al_{0.75}Ga_{0.25}N$ and lower by 0.65 (0.79) eV than in $Al_{0.85}Ga_{0.15}N$, thus the barrier for electron tunneling is lower by the same amount. Further progress in compositionally graded AlGaN contacts[57,58] is necessary for Al-rich random AlGaN alloys to be technologically viable. Related to the same problem, the small conduction band offset between Al-rich random AlGaN alloys and dielectrics, *e.g.*, AlN, can lead to large leakage currents. For example, the band offset is only 0.58 eV in the $Al_{0.75}Ga_{0.25}N$/AlN system, and 0.44 eV in $Al_{0.8}Ga_{0.2}N$/AlN. In contrast, the band offset is 1.0 eV between the 1ML superlattice and AlN, and 1.15 eV between the 2ML superlattice and AlN. The more favorable conduction band position of the superlattices compared to random AlGaN alloys results in better integration with dielectrics. Hence, lower specific contact resistance and better integration with dielectrics is made possible for the atomically thin superlattices thanks to their lower effective composition and lower conduction-band position compared to Al-rich random AlGaN alloys.

Although we have considered infinitely repeating periodic superlattices in this work, the structures that we have proposed can be experimentally realized by growing superlattices that are sufficiently thick. The electron thermal wavelength $\lambda_{th} = \sqrt{2\pi\hbar^2/m^*k_BT}$ is approximately 10 nm in AlGaN, and the scattering mean-free path $\lambda_{mfp} = \sqrt{3k_BT/m^*}\langle\tau\rangle$, which we estimated from our mobility calculations, is between 10 nm and 15 nm. For vertical transport, the superlattice stack thickness should exceed these length scales, with thicker stacks enabling higher breakdown voltages. For in-plane transport, we expect 30-nm-thick stacks to be sufficient, which is the typical thickness used for GaN quantum-well high-electron mobility transistors[39]. In terms of growth, thermodynamic mixing may occur at high growth temperatures between the AlN and GaN sublattice layers, thereby producing ternary $Al_xGa_{1-x}N$/$Al_yGa_{1-y}N$ superlattices, with $x \simeq 1$ and $y \simeq 0$. Since the mobility of Al-rich and Ga-rich AlGaN alloys is phonon-limited rather than disorder-limited, we expect the performance of the atomically thin superlattices to be robust against minor ternary-cation mixing in the sublattice layers. Therefore, the superlattices that we have proposed should be experimentally feasible as long as good uniformity in the sublattice composition and thickness is maintained.

Overall, the 1ML AlN/GaN superlattice has the largest modified BFOM among all known semiconductors with experimentally demonstrated dopability. Its modified BFOM is larger than the modified BFOM of $\beta$-$Ga_2O_3$ by a factor of ~3, 4H-SiC by a factor of ~7, cBN by a factor of ~12, Si by a factor of ~1300, and diamond by a factor of ~10,000. Table 4 lists the band gap, breakdown field, dielectric constant, dopant ionization energy, and carrier mobility that we used for the calculation of the BFOM and the modified BFOM for all semiconductors that we considered. We assume a carrier density of $10^{18}$ cm$^{-3}$ for all materials when calculating the dopant-ionization fraction. For consistency in our comparison of the modified BFOM with other materials, we used first-principles band gaps and phonon-limited mobilities calculated with many-body corrections and the iterative Boltzmann transport equation with dipole-corrected *ab-initio* electron-phonon matrix elements[10,43,59]; if not available, we used values that are widely accepted in the literature[5,56,60–62]. We calculated the breakdown fields using the model presented above, and obtained the dopant ionization energies from literature[5,18,63]. In addition to the 1ML superlattice, Al-rich AlGaN with Al composition below ~85% shows great promise for high-power devices if the technological challenges associated with high specific contact resistance and integration with non-native dielectrics, *e.g.*, MgO[64], can be resolved. However, these challenges are fundamentally related to the unfavorable position of their conduction bands, and the extent to which progress can be made is unclear. In this regard, the superlattices offer a clear advantage since they have a lower effective composition and lower conduction band, which allows for better integration with metals and dielectrics. Random AlGaN alloys require a minimum Al composition of 61% for their modified BFOM to be competitive with their closest non-nitride competitor, $\beta$-$Ga_2O_3$, which exhibits a modified BFOM of 3.7 GW/cm$^2$. The 2ML AlN/GaN superlattice also exhibits a high modified BFOM of 4.3 GW/cm$^2$ that is comparable to the modified BFOM of $\beta$-$Ga_2O_3$. Unlike $\beta$-$Ga_2O_3$, which suffers from severe self-heating due to low thermal conductivity (~20 W m$^{-1}$ K$^{-1}$)[55], III-nitride semiconductors have higher thermal conductivity (~200-300 W m$^{-1}$ K$^{-1}$ for ordered compounds[65–67]) thanks to weaker anharmonic phonon-phonon coupling. This enables efficient cooling and, therefore, high performance since the phonon-limited mobility decreases sharply with temperature. Finally, an advantage of the III-nitrides is that they are among the few ultra-wide-band-gap semiconductors for which both n-type and p-type doping has been experimentally demonstrated, which is necessary for ambipolar high-power devices[68].

Table 4. Comparison of the Baliga Figure of Merit and Modified Baliga Figure of Merit for various semiconductors. The monolayer-thin AlN/GaN digital-alloy superlattice surpasses all known ultra-wide-band-gap semiconductors for power-electronics applications.

| Material | $\varepsilon_G$ [eV] | $F_{br}$ [MV/cm] | $\epsilon_S$ | $\varepsilon_D$ [meV] | $\mu$ [cm$^2$/Vs] | BFOM [GW/cm$^2$] | MBFOM [GW/cm$^2$] |
|---|---|---|---|---|---|---|---|
| 1 ML GaN / 1 ML AlN Superlattice (this work) | 4.8 | 6.2 | 9.2 | 15[a] | 452 (∥) 369 (⊥) | 22 (∥) 18 (⊥) | **11.4 (∥)** **9.3 (⊥)** |
| 2 ML GaN / 2 ML AlN Superlattice (this work) | 4.6 | 5.7 | 9.2 | 15[a] | 210 | 8.0 | **4.3** |
| Random Al$_{0.75}$Ga$_{0.25}$N (this work) | 5.5 | 8.1 | 8.8 | 18[a] | 125 | 13 | **6.4** |
| Random Al$_{0.5}$Ga$_{0.5}$N (this work) | 4.8 | 6.2 | 9.1 | 15[a] | 115 | 5.6 | **2.6** |
| AlN (this work) | 6.3 | 11 | 8.5 | 255[a] | 373 | 87 | **3.5×10$^{-3}$** |
| $\beta$-Ga$_2$O$_3$ | 4.8[b] | 6.2 | 10[b] | 30[b] | 200[b] | 11 | **3.7** |
| GaN | 3.5 | 3.8 | 9.7 | 15[a] | 830[c] | 6.4 | **2.2** |
| 4H-SiC | 3.2[a] | 3.1 | 9.7[b] | 60[b] | 900[b] | 4.1 | **1.7** |
| cBN | 6.8[d] | 12 | 7.1[d] | 250[d] | 1610[d] | 490 | **0.95** |
| Si | 1.1[b] | 0.3 | 11.7[b] | 45[b] | 1400[e] | 1.2×10$^{-2}$ | **8.8×10$^{-3}$** |
| Diamond | 5.7[d] | 8.8 | 5.7[d] | 370[d] | 1970[d] | 170 | **1.1×10$^{-3}$** |

[a]Ref.[18] [b]Ref.[5] [c]Ref.[10] [d]Ref.[59] [e]Ref.[43]

In summary, we propose an experimentally feasible design, *i.e.*, atomically thin superlattices of AlN and GaN, that removes alloy scattering in AlGaN and, therefore, enhances its power-electronics figure of merit. Our calculations show that AlN/GaN superlattices are promising semiconductors for next-generation power electronics due to their ultra-wide band gap, high electron mobility, and availability of shallow dopants. They exhibit the largest modified BFOM among all the technologically relevant semiconductors that we have considered. Moreover, such superlattices offer lower specific contact resistance and better integration with dielectrics compared to Al-rich random AlGaN alloys. Most importantly, similar superlattices have already been demonstrated experimentally using industrial growth techniques. Similar theoretical characterization and materials prediction from first principles will enable the discovery of efficient semiconductors for a wide range of device applications.

See the Supplementary Material for (1) details of our mobility calculations, (2) a comparison of the phenomenological model for the breakdown field against experiments, (3) the ionization energy of Si as a function of Al composition, and (4) the composition-dependent conduction band offset of the atomically thin superlattices and random AlGaN alloys.

We thank Josh Leveillee for helpful discussions. This work was supported as part of the Computational Materials Sciences Program funded by the U.S. Department of Energy, Office of Science, Basic Energy Sciences, under Award No. DE-SC0020129. Computational resources were provided by the National Energy Research Scientific Computing (NERSC) Center, a Department of Energy Office of Science User Facility supported under Contract No. DEAC0205CH11231. N. Pant acknowledges the support of the Natural Sciences and Engineering Research Council of Canada Postgraduate Doctoral Scholarship.

*Supplementary Material*
Increasing the Mobility and Power-Electronics Figure of Merit of AlGaN with Atomically Thin AlN/GaN Superlattices

Nick Pant[1,2], Woncheol Lee[3], Nocona Sanders[2], and Emmanouil Kioupakis[2]

1. Applied Physics Program, University of Michigan, Ann Arbor
2. Department of Materials Science and Engineering, University of Michigan, Ann Arbor
3. Department of Electrical Engineering and Computer Science, University of Michigan, Ann Arbor

**Details of Mobility Calculations**

The EPW code uses the following definition of the low-field mobility [1],

$$\mu_{\alpha\beta} = -\frac{1}{V_{PC} n_c} \sum_n \int \frac{d^3 \mathbf{k}}{\Omega_{BZ}} v_{n\mathbf{k},\alpha} \partial_{E_\beta} f_{n\mathbf{k}} \quad (S1),$$

where $\alpha, \beta$ are cartesian coordinates, $V_{PC}$ is the volume of the primitive cell, $n_c$ is the carrier density, $n$ is the band index, $\mathbf{k}$ is a crystal wave vector in the first Brillouin zone, and $\Omega_{BZ}$ is the volume of the first Brillouin zone. The band velocity $v_{n\mathbf{k},\alpha}$ corresponds to the momentum-space gradient of the bands, $v_{n\mathbf{k},\alpha} = \frac{1}{\hbar} \nabla_{\mathbf{k},\alpha} \varepsilon_{n\mathbf{k}}$ [1]. The quantity $\partial_{E_\beta} f_{n\mathbf{k}}$ is the linear response of the electronic occupation function to a small electric field $E$ applied along the $\beta$ direction, which we obtained by self-consistently solving the linearized Boltzmann transport equation [1],

$$\partial_{E_\beta} f_{n\mathbf{k}} = e v_{n\mathbf{k}\beta} \frac{\partial f_{n\mathbf{k}}^0}{\partial \varepsilon_{n\mathbf{k}}} \tau_{n\mathbf{k}}$$
$$+ \frac{2\pi \tau_{n\mathbf{k}}}{\hbar} \sum_{mv} \int \frac{d^3 \mathbf{q}}{\Omega_{BZ}} |g_{nmv}(\mathbf{k},\mathbf{q})|^2$$
$$\times [(n_{v\mathbf{q}} + 1 - f_{m\mathbf{k}+\mathbf{q}}^0) \delta(\varepsilon_{n\mathbf{k}} - \varepsilon_{m\mathbf{k}+\mathbf{q}} - \hbar\omega_{v\mathbf{q}})$$
$$+ (n_{v\mathbf{q}} + f_{m\mathbf{k}+\mathbf{q}}^0) \delta(\varepsilon_{n\mathbf{k}} - \varepsilon_{m\mathbf{k}+\mathbf{q}} + \hbar\omega_{v\mathbf{q}})] \partial_{E_\beta} f_{m\mathbf{k}+\mathbf{q}} \quad (S2),$$

$$\frac{1}{\tau_{n\mathbf{k}}} = \frac{2\pi}{\hbar} \sum_{mv} \int \frac{d^3 \mathbf{q}}{\Omega_{BZ}} |g_{nmv}(\mathbf{k},\mathbf{q})|^2$$
$$\times [(n_{v\mathbf{q}} + 1 - f_{m\mathbf{k}+\mathbf{q}}^0) \delta(\varepsilon_{n\mathbf{k}} - \varepsilon_{m\mathbf{k}+\mathbf{q}} - \hbar\omega_{v\mathbf{q}})$$
$$+ (n_{v\mathbf{q}} + f_{m\mathbf{k}+\mathbf{q}}^0) \delta(\varepsilon_{n\mathbf{k}} - \varepsilon_{m\mathbf{k}+\mathbf{q}} + \hbar\omega_{v\mathbf{q}})] \quad (S3).$$

In the equations above, the quantum numbers $n$ and $m$ are electronic band indices, $\mathbf{k}$ is the electronic crystal wave vector, $v$ is the phonon branch index, and $\mathbf{q}$ is the phonon wave vector in the first Brillouin zone. We calculated the electronic eigenvalues $\varepsilon_{n\mathbf{k}}$ in the G$_0$W$_0$ approximation, from which the Fermi-Dirac occupation factors $f_{n\mathbf{k}}^0$ are calculated at room temperature. For the mobility calculation, we included electronic states within 300 meV of the conduction-band edge since higher states do not contribute to low-field transport due to their small occupation. We calculated the phonon eigenvalues $\omega_{v\mathbf{q}}$ from density-functional perturbation theory in the local-density approximation, from which the Bose-Einstein occupation factors $n_{v\mathbf{q}}$ are calculated. Density-functional perturbation theory also produces the electron-phonon matrix elements $g_{nmv}(\mathbf{k},\mathbf{q})$, which give the probability amplitudes for the interband and intraband scattering processes from electronic states $|n\mathbf{k}\rangle$ to $|m\mathbf{k}+\mathbf{q}\rangle$ mediated by all $3\times N_{atom}$ phonon modes $|\nu\mathbf{q}\rangle$ throughout the Brillouin zone. The integrals over $\mathbf{k}$ and $\mathbf{q}$ in equations (1), (2), and (3) converge for very fine grids with $O(100^3)$ grid points, however density-functional and density-functional perturbation theory calculations are typically performed for coarser grids with $O(10^3)$ grid points due to their computational cost. This challenge is overcome by interpolating the coarse-grid electronic and phononic eigenvalues, velocity matrix elements, and electron-phonon scattering matrix elements to fine grids using the maximally localized Wannier-function method as implemented in the EPW code[2,3]. In polar materials, the electron-phonon matrix elements of longitudinal-optical Fröhlich modes exhibit a $O(1/q)$ divergence as $q \to 0$, due to the dipole charge contribution to the electron-phonon interaction[4]. This presents a challenge for Wannier interpolation since divergent functions do not have well-behaved Fourier transforms. We overcame this challenge using the EPW code by analytically treating the long-range dipolar divergence while numerically treating the well-behaved short-range interaction[4]. Overall, we interpolated the necessary quantities to 160×160×110 fine $\mathbf{k}$ and $\mathbf{q}$ grids for primitive-cell structures, *i.e.*, GaN, AlN, and the 1ML AlN/GaN superlattice, and 160×160×55 fine $\mathbf{k}$ and $\mathbf{q}$ grids for the 2ML AlN/GaN supercell structure to obtain converged phonon-limited mobilities.

Unlike electron-phonon scattering, electron-alloy scattering is elastic and has no angular dependence, thus the relaxation-time approximation can be more reliably used to calculate the alloy-scattering mobility. We calculated the alloy-scattering-limited mobility using an in-house code in the relaxation-time approximation, where the composition-dependent relaxation time is given by the formula,

$$\frac{1}{\tau(\varepsilon)} = \frac{2\pi}{\hbar} U_0^2\, x(1-x) V_{PC} \frac{m^{*3/2}}{\sqrt{2}\pi^2 \hbar^3} \sqrt{\varepsilon} \quad (S4),$$

where $x$ is the aluminum composition. In this equation, we used the geometrically averaged conduction-band effective mass $m^*(x)$, which we calculated for random alloys as a linear interpolation of the G$_0$W$_0$ effective masses of GaN and AlN. In our previous work[5], we found that the alloy-scattering potential $U_0(x)$ in AlGaN can be calculated by unfolding the band structure of special-quasirandom-structure supercells and fitting the alloy-scattering-rate expression to the energy-broadened effective band structure. In the same work, we showed that a one-to-one correspondence exists between the alloy-scattering potential $U_0(x)$ and the slope of the conduction band versus composition curve $\partial \varepsilon_c / \partial x$, obtained by referencing the branch-point energies of the alloys. For materials where the conduction band is a quadratic function of composition, *e.g.*, AlGaN, it turns out that the slope $\partial \varepsilon_c / \partial x$, and therefore the scattering potential, at $x=0.5$ is equal to the conduction band offset $\Delta \varepsilon_C$ between the end binary compounds. By referencing the branch-point energies of GaN and AlN, we calculated a conduction-band offset $\Delta \varepsilon_C \approx 1.8$ eV at the G$_0$W$_0$ level, which is equal to the alloy-scattering potential $U_0(x=0.5) = 1.8$ eV that we calculated in our previous work by unfolding the band structure at the LDA and hybrid-functional level. We excluded neutral defect and ionized impurity scattering in our calculation of the room-temperature mobility, therefore our mobility estimates serve as theoretical upper bounds.

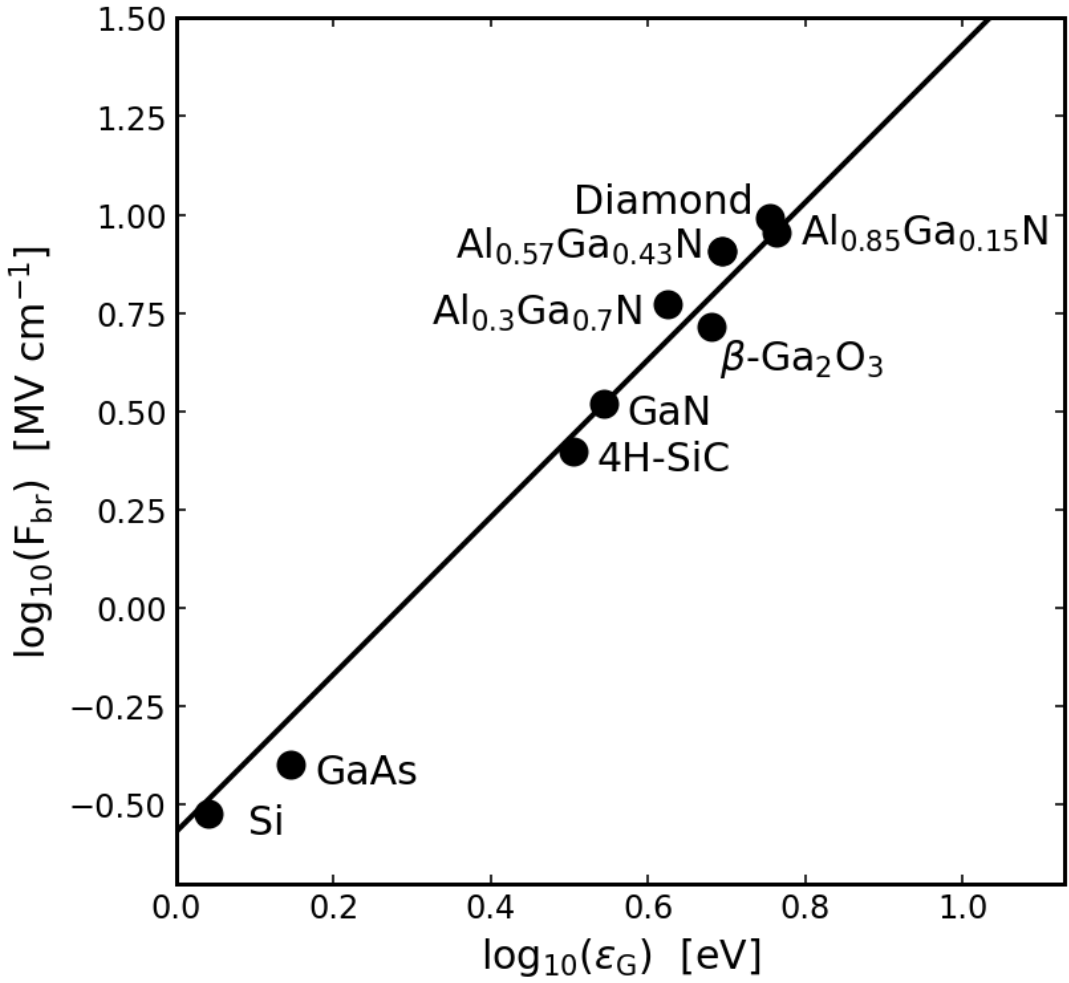

Figure S1. Breakdown field as a function of the band gap in a wide range of semiconductor. The scatter points are experimental values [6–12], and the solid line is the phenomenological model, $F_{br} = 3.3 \text{ MV/cm} \times (\varepsilon_G/3.5)^2$.

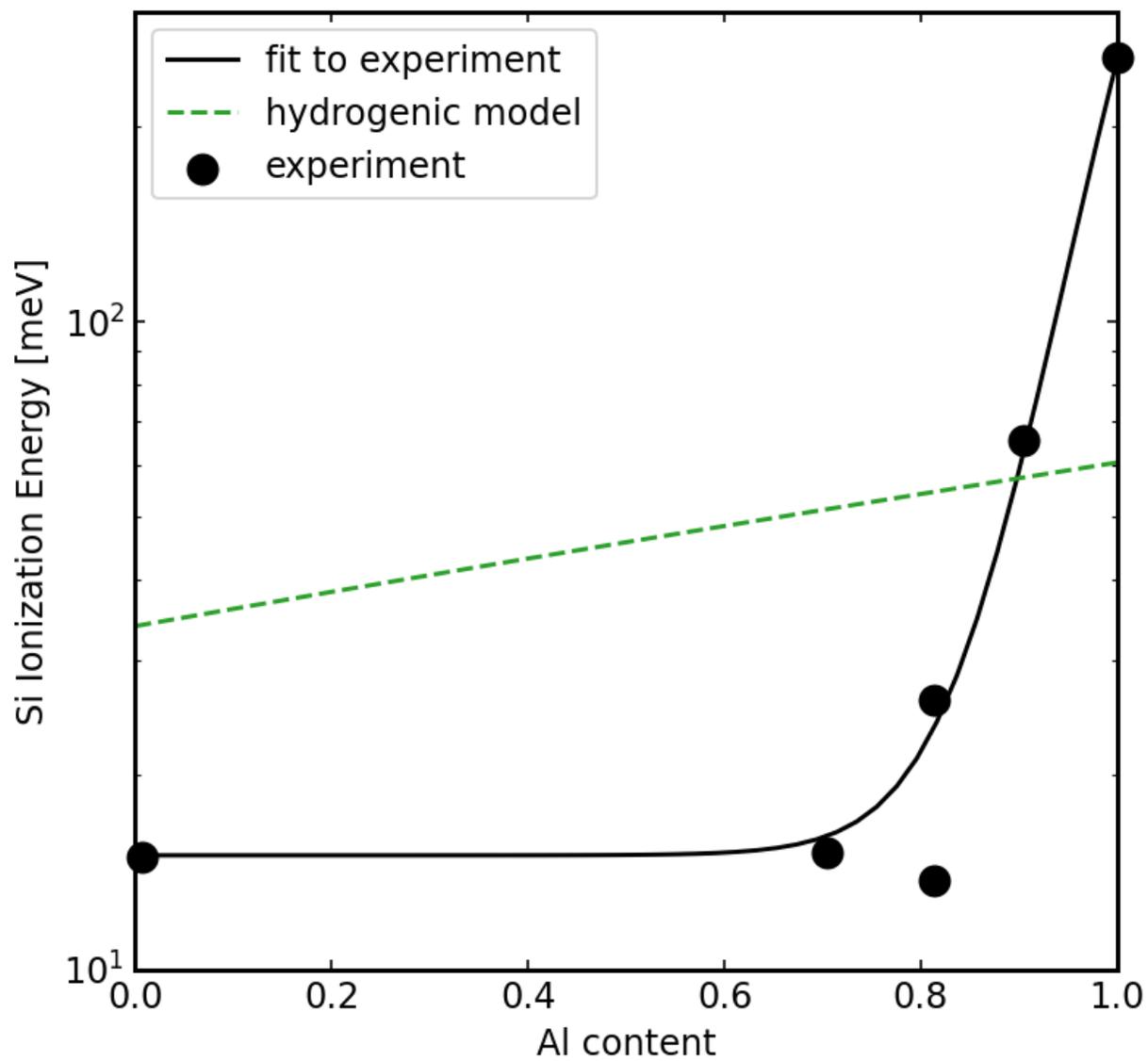

Figure S2. The ionization energy of Si in AlGaN as a function of Al composition. The experimental data points are obtained from Collazo *et al.* [18]

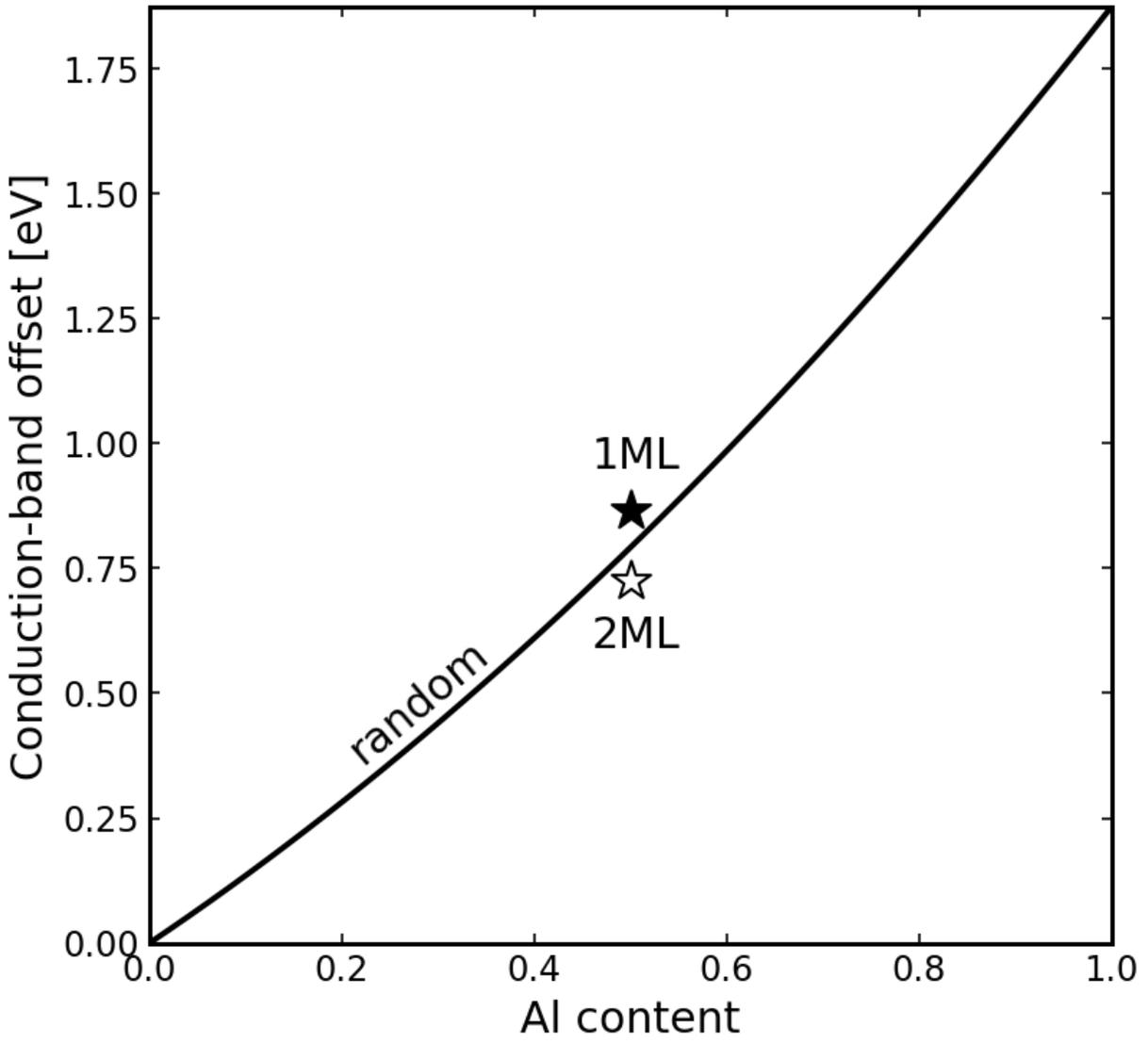

Figure S3. Conduction-band offset of random AlGaN alloys (solid curve) and atomically thin superlattices (stars) as a function of Al composition. The band offset is given relative to the conduction-band position of GaN, which we evaluated by referencing their branch-point energies. For random AlGaN alloys, we used the bowing parameter for the conduction band calculated by Kyrtsos *et al.* [14]